\documentclass[preprint,12pt]{elsarticle}

\usepackage{amssymb}
\usepackage{subfigure}
\usepackage{graphicx}
\usepackage{aas_macros}

\biboptions{comma,square}

\journal{Nuclear Instruments and Methods A}

\begin{document}

\begin{frontmatter}


\ead{paolo.soffitta@iaps.inaf.it}
\fntext[label2]{Corresponding Author}


\title{Measurement of the position resolution \\of the Gas Pixel Detector}


\author{Paolo Soffitta\corref{cor1}\fnref{label2}} \author{Fabio Muleri, Sergio Fabiani, Enrico Costa}

\address{IAPS/INAF Via Fosso del Cavaliere 100, 00133-Roma, Italy}

\author{Ronaldo Bellazzini, Alessandro Brez, Massimo Minuti, \\Michele Pinchera, Gloria Spandre}

\address{INFN-Pisa Largo B. Pontecorvo, 3 - 56127 Pisa, Italy}

\begin{abstract}
The Gas Pixel Detector was designed and built as a focal plane instrument for X-ray polarimetry of celestial sources, the
last unexplored subtopics of X-ray astronomy.
It promises to perform detailed and sensitive measurements resolving extended sources and detecting polarization in faint sources
in crowded fields at the focus of telescopes of good angular resolution. Its polarimetric and spectral capability
were already studied in earlier works. Here we investigate for the first time, with both laboratory measurements
and Monte Carlo simulations, its imaging properties to confirm its unique capability to carry out imaging spectral-polarimetry
in future X-ray missions.

\end{abstract}

\begin{keyword}

X-ray; astronomy; detectors; polarimetry
\end{keyword}

\end{frontmatter}


\section{Introduction}
\label{Intro}

The Gas Pixel Detector (GPD) \cite{2001Natur.411..662C,
2006NIMPA.566..552B, 2007NIMPA.579..853B} has been designed and built
to perform time resolved imaging spectral-polarimetry of X-ray celestial sources by means of
the photoelectric effect. It is the long sought quantum leap in sensitivity with respect to the classical
techniques \cite{2003NIMPA.510..170S}. Successively the same method was implemented
exploiting an alternative technique without any 2-D imaging capability \cite{2007NIMPA.581..755B}.

While the polarimetric and the spectroscopic performances of the
GPD have been already studied in detail \cite{2008NIMPA.584..149M,
2010NIMPA.620..285M}, its imaging properties have been
estimated, so far, only by Monte Carlo simulations evaluating the point of conversion of
the impinging photon with a suitable algorithm \cite{2003SPIE.4843..383B}. Still the GPD was devised
primarily to be exploited at the focus of conventional and multi-layer hard X-ray optics \cite{2006SPIE.6266E..85M}.
At this purpose the GPD has been considered for different proposed
space missions \cite{2008SPIE.7011E..62S,2010ExA....28..137C,2010xpnw.book..269B,2010SPIE.7732E..38S}.
While X-ray optics with the exquisite quality of those of Chandra will not be again available in the near future,
for POLARIX (or XIPE, proposed in 2012 to ESA as a small mission for a launch on 2017)
the JET-X optics (three complete mirrors, two Flight Models, FM, the heritage of the former
Spectrum X-Gamma mission, and one Qualification Model), can still be used. They have a Point Spread
Function (PSF) with a measured Half Energy Width (HEW) at 1.5-keV of 15.l'' (FM-1) and 14.6'' (FM-2)
\cite{1997SPIE.3114..392W}.

We report on the first measurement, with a narrow beam at 4.5 keV, of the
position resolution of a GPD filled with a mixture of He-DME 80-20 (1 bar) with a large Gas Electron Multiplier (GEM),
This proves the capability of the GPD to perform space-resolved measurements, such as those of Pulsar-Wind Nebulae, Supernova Remnants or
X-ray jets with a suitable accuracy, and to observe the faintest source accessible to X-ray polarimetry especially in crowded fields \cite{2012SPIE.8443S }.
We also compared such measurement with the Monte Carlo simulation relating the
characteristics of the image to the modulation factor. Finally we showed the effect on imaging
of the non-uniformity of the drift electric field in case the GEM plane, and in particularly
the guard-ring, is not powered.

\section{The GPD prototype and the relevant algorithms}
\label{sec: GPD&algorithms}
The GPD is a gas cell with a beryllium thin window, a drift region, a
perforated multiplication plane (the GEM) and a collecting plane with an
hexagonal pattern.

The collecting (or readout) plane is the top layer of an ASIC CMOS with a square surface of
1.5 cm $\times$ 1.5 cm having 105600 hexagonal pads of 50 $\mu$m pitch.
It is designed with a very high precision due to the micro-electronics
technologies employed. The positions of the hexagonal pads set up, therefore, a very high accurate
reference system which is essential for an accurate imaging.

Hereafter we summarize, to understand the considerations in the following sections, the sequential steps of the algorithm that processes the
signal from each hexagonal pad that collects the charges of the track. The algorithm is aimed to evaluate:

\begin{itemize}
    \item the \emph{barycentre} using all the pixels;
    \item the \emph{angle}, for which the second moment, with respect to the barycentre, is maximum;
    \item the \emph{third moment} (skewness), also with respect to the barycentre, to determine which of the two end points contains the impact point
          (by the smaller charge density);
    \item  the \emph{impact point}, using the pixel only within a sub-region of the initial part of the track as it is identified by the skewness.
                The extension of this sub-region depends on the gas mixture and it is evaluated by using the maximum second moment as a weight;
    \item the \emph{angle} (emission direction) that maximizes the second moment with respect to the impact point weighting the charges according to
           their distance from it.

\end{itemize}

\begin{figure}[htpb]
\centering
\subfigure[\label{GPD_LEP_OLD}]{\includegraphics[scale=0.2]{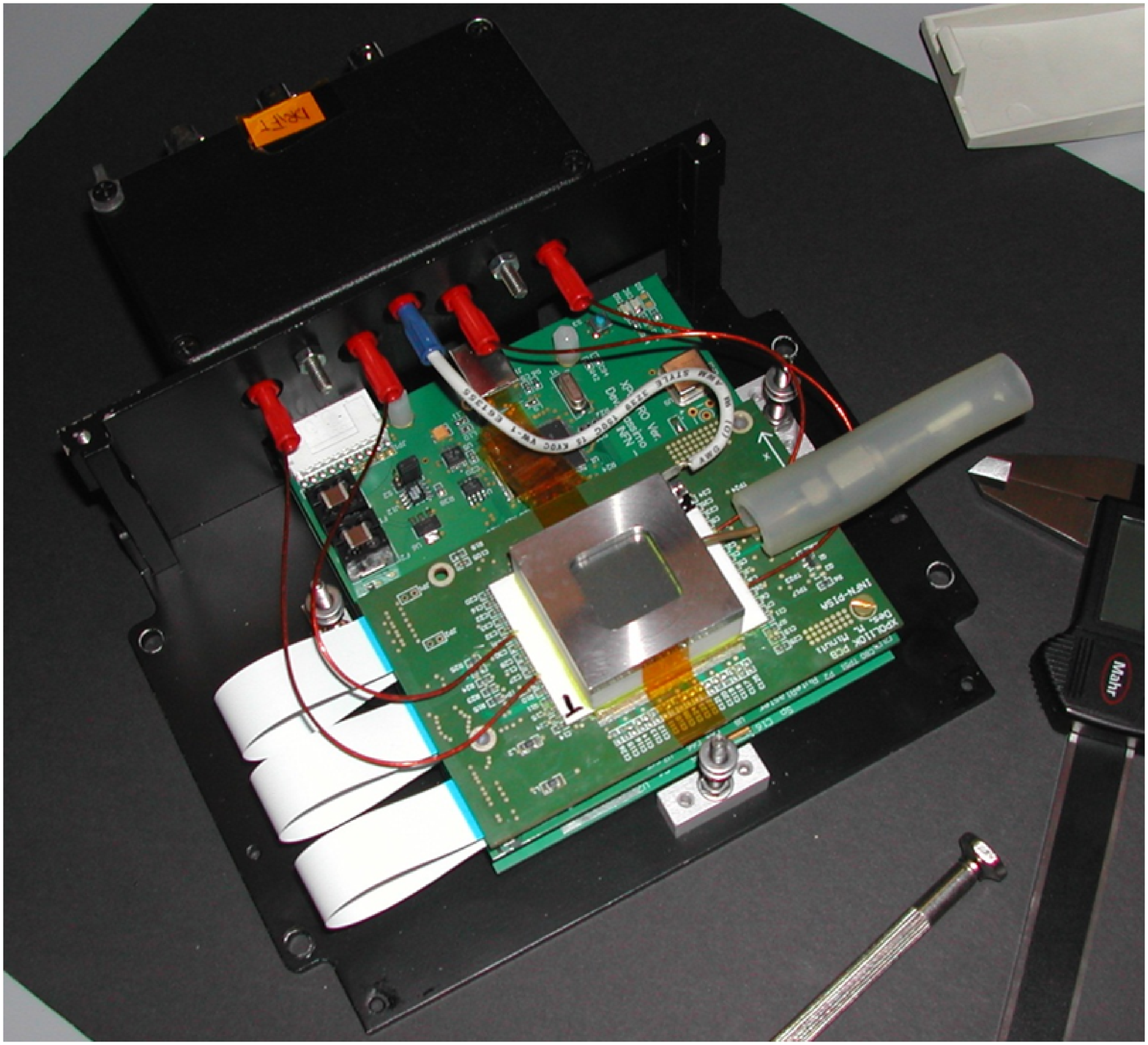}}
\hspace{1cm}
\subfigure[\label{GPD_LEP_NEW}]{\includegraphics[scale=0.2]{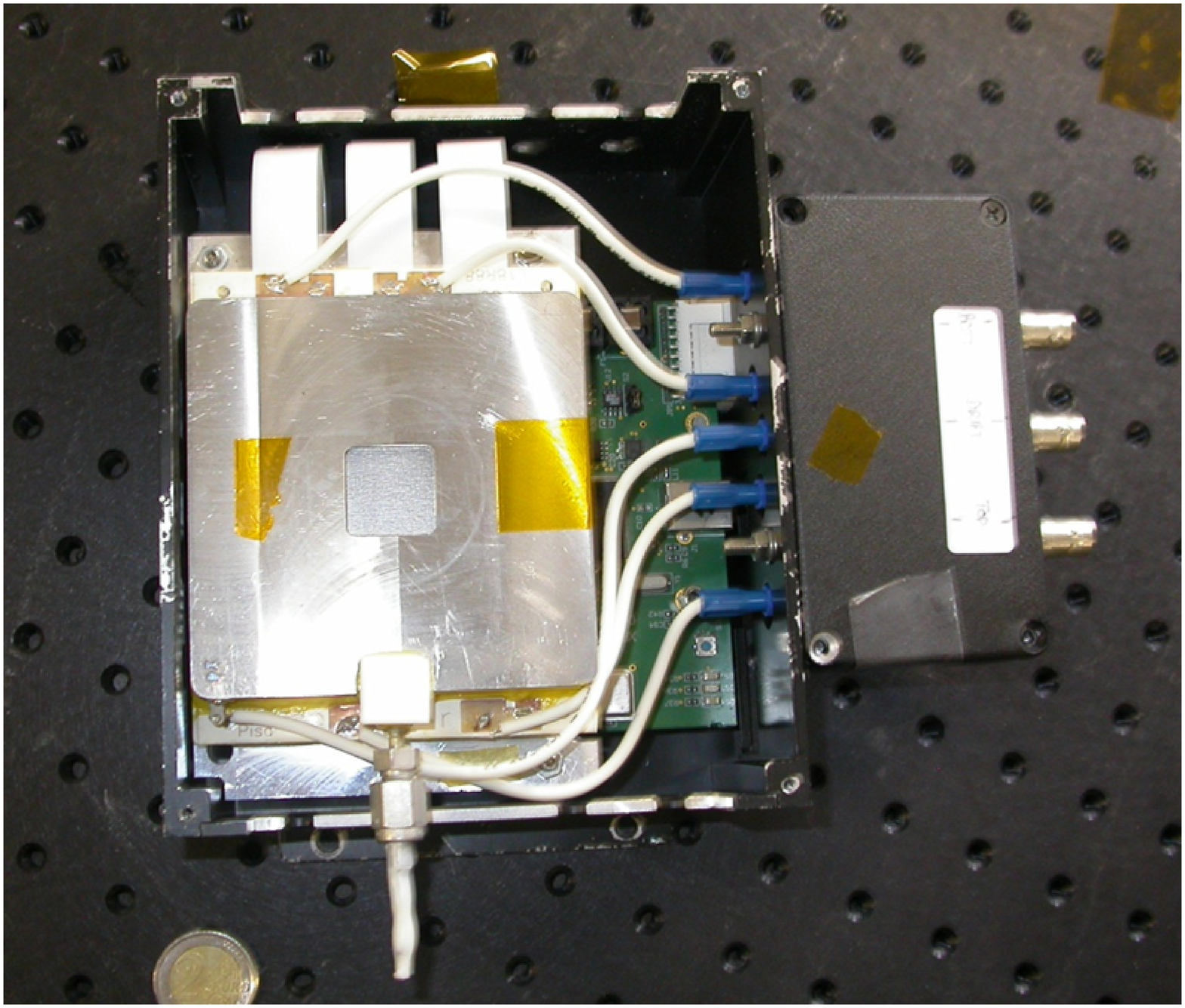}}
\caption{({\bf a}). The old design of the GPD with a narrow guard-ring in the GEM plane ({\bf b}). The present design with a larger guard-ring \cite{2012SPIE.8443 }.}
\end{figure}

The first sealed GPD \cite{2007NIMPA.579..853B} had a GEM plane with the perforated
surface as large as the  ASIC CMOS chip and a
guard-ring, not perforated, just slightly larger. The present
prototype \cite{2012SPIE.8443 }  has, instead, a much
larger guard-ring extending as far as 2.25 cm  from the edge of the
ASIC chip. This latter design, see fig. \ref{GPD_LEP_OLD}  and fig. \ref{GPD_LEP_NEW} for a comparison,
allows for a better control of the
uniformity of the electric field.  Border effects, other than those
due to the extension of the track and to its incomplete imaging close to the edge
of the sensitive region, are now therefore expected to be vanishing.
We will show later how the guard-ring influences the position resolution
of the GPD.

The GPD  has a drift/absorption length of 1 cm and a beryllium window 50 $\mu$m thick.
The GEM employed with this new design is 50 $\mu$m thick, with holes of 30 $\mu$m of diameter with 50 $\mu$m pitch.
The holes are laser etched on a liquid crystal polymer substrate. The GEM is manufactured by SciEnergy under the supervision
of  RIKEN \cite{2010xpnw.book...60T}.

\section{The X-ray beam profile}
\label{sec: Beam_profile}
In order to measure the position resolution we arranged a narrow X-ray beam by means of
a collimator system connected to the exit beryllium window of a 50 W
(50 kV, 1 mA) commercial X-ray tube with a Titanium anode by Oxford Instruments
(series 5000) having an anode spot smaller than 150 $\mu$m.
The X-rays from the tube, operated at 30 kV and
400 $\mu$A, were collimated by means of two diaphragms, one made
of brass having a diameter of 500 $\mu$m connected to the exit window of the
X-ray tube and one made of tungsten with a diameter of 25 $\mu$m on the opposite side close
to the detector window. The two diaphragms were interconnected by two coaxial brass shielding tubes and
separated by 17 cm.
The large distance necessary to reduce the beam divergency does not allow
for measuring the position resolution at energies below the Titanium X-ray K
lines at 4.51 keV $(K_{\alpha})$ due to the intervening air absorption.
In order to measure the actual beam size, we scanned the X-ray beam with a
brass slit with sharp edges, at the same distance of the GPD, in two orthogonal directions
and at a constant speed using as detector a commercial one-pixel
Si-PIN XR-100-CR manufactured by Amptek. The geometrical
surface of the detector is 5 mm $\times$ 5 mm.
This system, when interfaced to the two orthogonal linear stages with 0.5
$\mu$m resolution (model ILS50CC), manufactured by Newport and operated via
computer by means of an 8-axis XPS controller \cite{2008SPIE.7011E..61M}, allows for
continuously shifting the detector and the slit with respect to
the beam at a minimum speed of 1 $\mu$m $s^{-1}$. Data from the
Si-PIN detector were acquired in rate-meter mode with a time-bin
of 1 s, selecting a window in energy to include only the titanium
K-lines. With this set-up, the time needed for obscuring the
beam at the slit passage  provides the beam size along the
direction of the scan.

If the beam is described by a 2-D gaussian, as in the following equation \ref{bidimgauss}) :

\begin{equation}
\label{bidimgauss}
    f(X,Y) = A e^{- (\frac{(X-X_\circ)^{2}}{2 \sigma_{X}^{2}} + \frac{(Y-Y_\circ)^{2}}{2 \sigma_{Y}^{2}})}
\end{equation}

the counting rate as a function of time (or as a function of position being the speed constant and known) resulting from the scanning along the X
(or Y) direction can be described by the \emph{erf} function as in eq. \ref{erf}. This is, actually, the primitive of eq.
 \ref{bidimgauss} after the integration between -$\infty$ and +$\infty$ of the counts in Y (or X) direction.

\begin{figure}[htpb]
\centering
\subfigure[\label{fig: scanY}]{\includegraphics[scale=0.35]{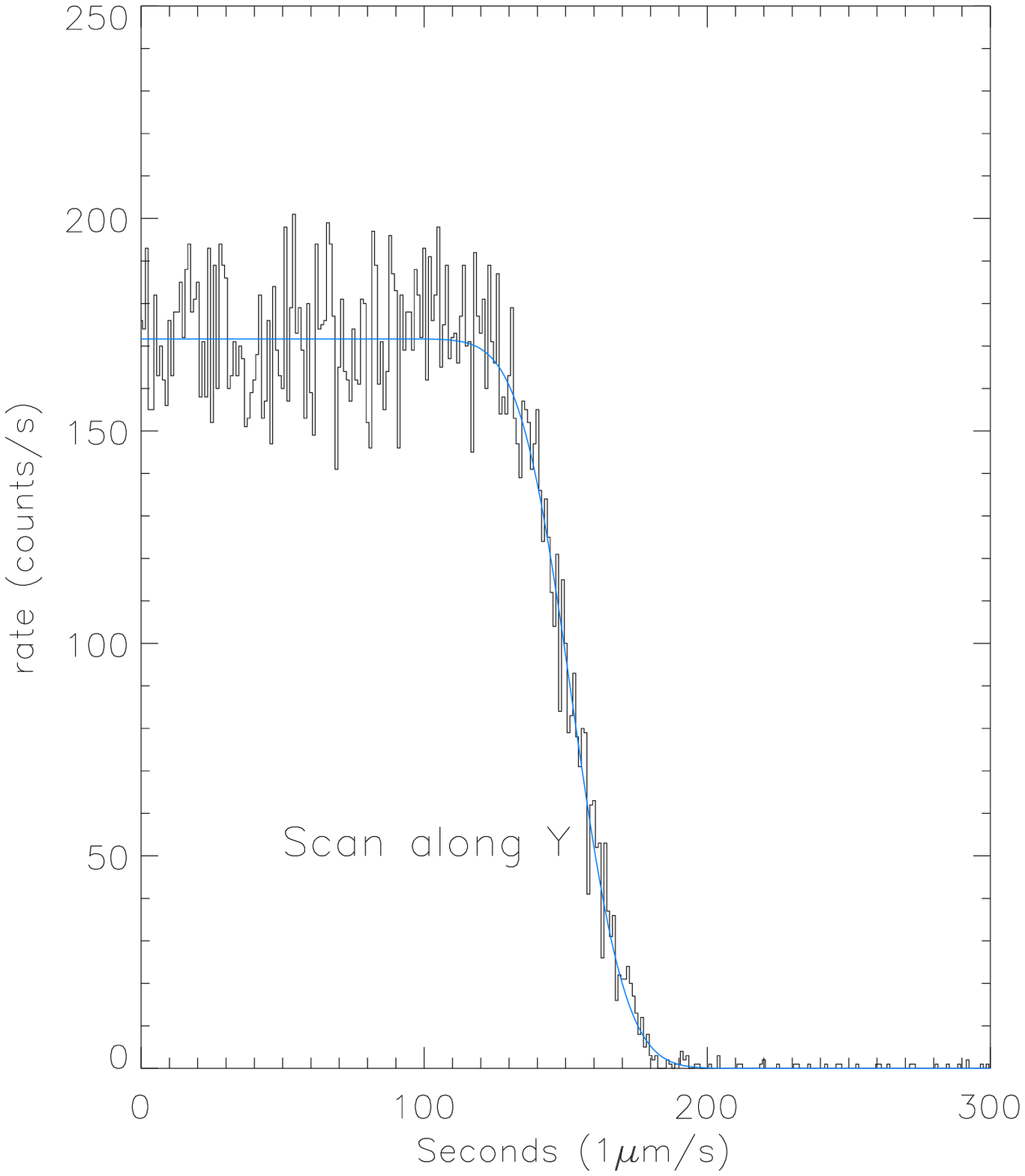}}
\hspace{1cm}
\subfigure[\label{fig: scanX}]{\includegraphics[scale=0.35]{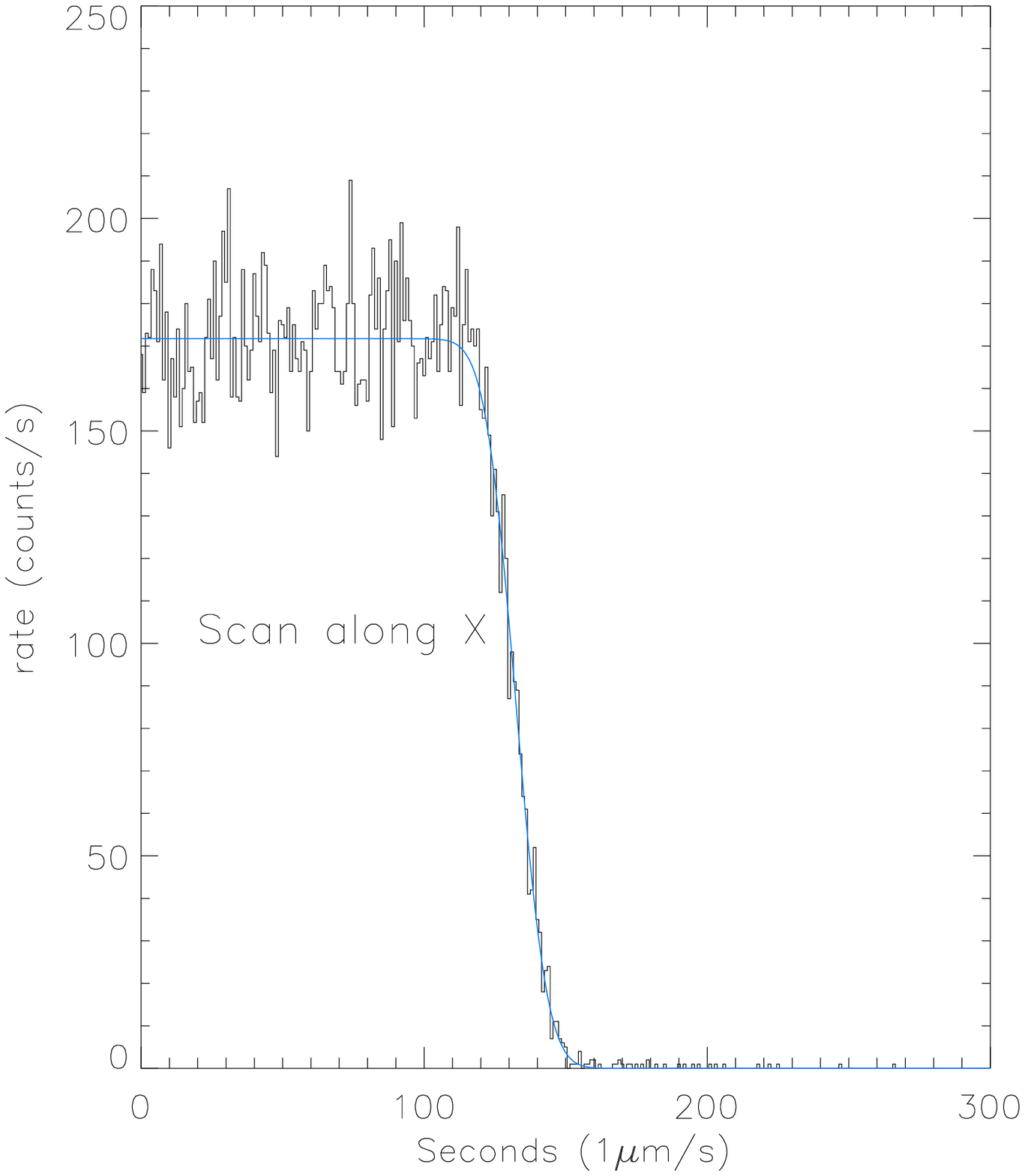}}
\caption{({\bf a}). Scan along the Y axis. The result from fitting eq \ref{erf} is $\sigma_{Y}$ = (14.7 $\pm$ 0.4) $\mu$m ({\bf b}). Scan along the X axis.
The result from fitting eq \ref{erf} is $\sigma_{X}$ = (8.7 $\pm$ 0.3) $\mu$m.}
\label{twobeamsizescan}
\end{figure}

The counting rate during the scanning in Y and X are shown in figure \ref{twobeamsizescan} together with the fitting function :

\begin{equation}
\label{erf}
    f(X) = \frac{A_{X}}{2}[1 - erf(\frac{\sqrt{2} (X-X_{0})}{2\sigma_{X}})]
\end{equation}

Such fitting procedure allows for deriving the relevant parameters ($\sigma_{X}$ and $\sigma_{Y}$) of the X-ray beam and comparing them with the results of
the section \ref{sec: pos res}. The results of the fit are shown in the table \ref{tab_res_beam_scan}:

\begin{table}[h]
\caption{Result of fitting procedure of the beam scan. The $\chi^{2}_{\nu}$ is 1.0 for the X scan and
1.12 for Y scan.}
\label{tab_res_beam_scan}
\begin{center}
\begin{tabular}{|c|c|}
  \hline
  Param & Fit results \\
  \hline
  $A_{X}$ & (171.71 $\pm$ 1.21) Counts/s \\
  $X_{0}$ & (132.42 $\pm$ 0.31) $\mu$m \\
  $\sigma_{X}$ & (8.7 $\pm$ 0.3) $\mu$m \\
  \hline
  $A_{Y}$ & (171.67 $\pm$ 1.16) Counts/s \\
  $Y_{0}$ & (152.41 $\pm$ 0.78) $\mu$m \\
  $\sigma_{Y}$ & (14.7 $\pm$ 0.4) $\mu$m \\
  \hline
\end{tabular}
\end{center}
\end{table}

The beam shape is therefore consistent with being gaussian with two different elongation
in the two orthogonal directions.
Such different elongations  may be due either to the
position of the 500 $\mu$m diaphragm with respect to the
anode spot of the X-ray tube or to a possible residual misalignment
between the two small diaphragms at the opposite end of the two collimator brass tubes.

We did not investigate further the cause of the two different sizes since their values are already smaller than the expected position resolution
of the GPD.

\section{The position resolution determination}
\label{sec: pos res}

To study the imaging capability of the GPD we
made a set of three measurements shifting it by 300 $\mu$m with respect to the central position (the center of the ASIC chip readout plane)
in orthogonal directions. This central position was, actually, found manually by iterating once the shift and the derived
centroid of the image of the beam.

\begin{figure}
\centering
\includegraphics [scale=0.5]{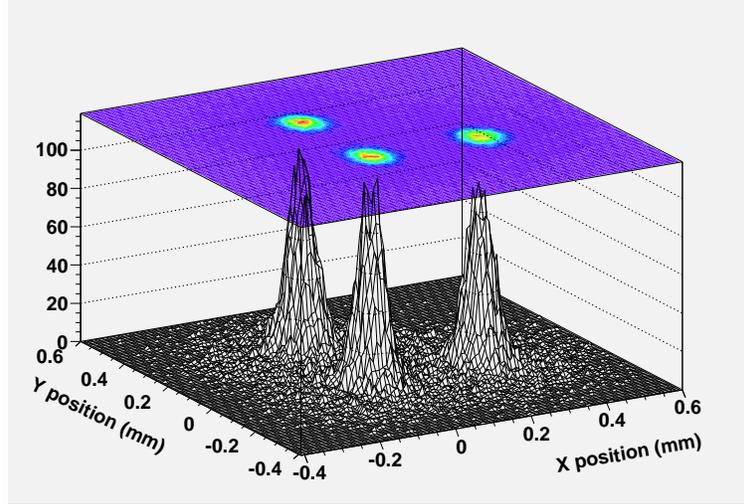}
\caption{The real images of the beam detected by the GPD. These images were
obtained by shifting the GPD by 300 $\mu$m in orthogonal directions with respect to the central position.}
\label{3_beams}       
\end{figure}

\begin{figure}[htpb]
\centering
\subfigure[\label{2076_paper_data}]{\includegraphics[scale=0.30]{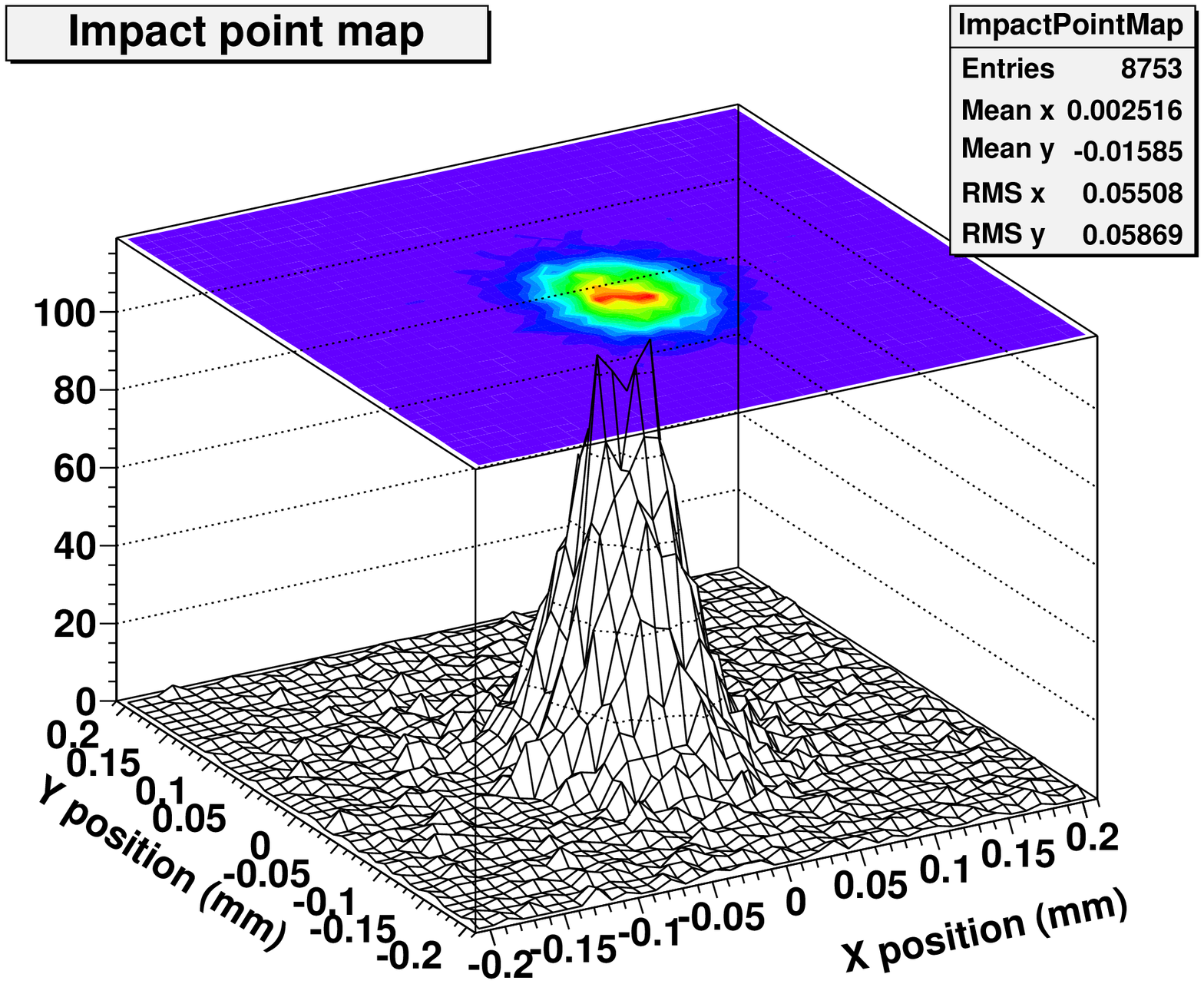}}
\hspace{0cm}
\subfigure[\label{2076_paper_fit}]{\includegraphics[scale=0.30]{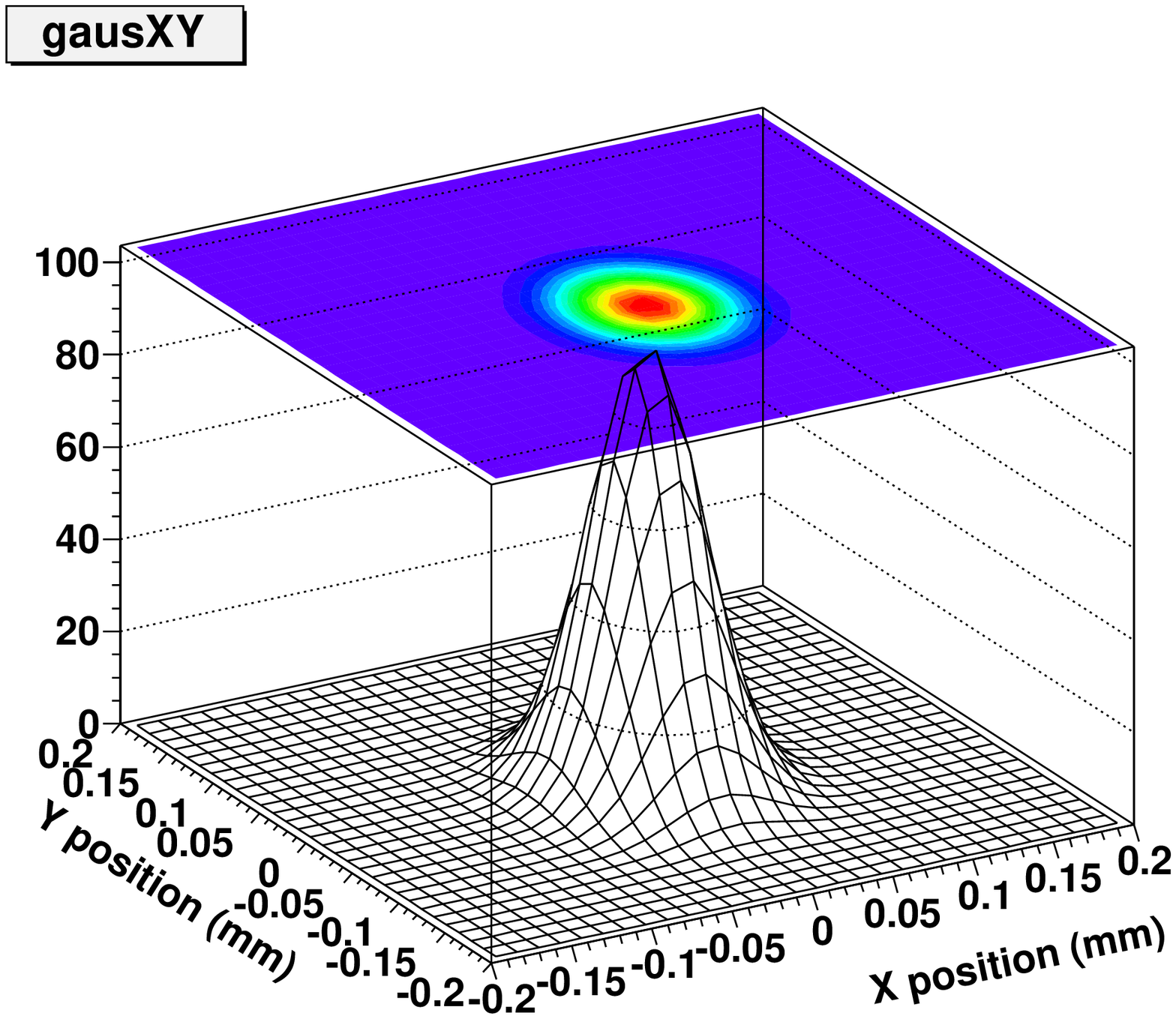}}
\caption{({\bf a}). A close-up of the image of the central beam. ({\bf b}). 2-D Gaussian function fitted to
the core of the image.}
\end{figure}

The counting rate during each acquisitions was about 10 counts $s^{-1}$, by
far larger than the background counting rate, for a total of about 8700 counts/acq.,
all windowed in energy to include only the Ti $K$ photo-peak.
The images of the three acquisitions all together are displayed in figure \ref{3_beams} where they appear well separated.
A closer view shows that they are characterized by a dominating core with external wings. We investigated the properties of the core and the wings
in detail later in section \ref{sec: Monte_Carlo} by means of Monte Carlo simulations.
In this section we, instead, measured the properties of the core. A bi-dimensional gaussian including two free
$\sigma_{X}$ and $\sigma_{X}$ parameters  (see eq.\ref{bidimgauss}) was fitted to the
data considering only the central part of the impact point distribution in a frame of
120 $\mu$m $\times$ 170 $\mu$m around the peak position. The fit was performed by using
the MINUIT package included in the ROOT 5.0 environment. The results of the fit are shown
in the table \ref{tab_res_fit}. The $\chi^{2}$ of the fit of the core was 200.9 with 196
degree of freedom. A close-up view of the image of beam in the central position and the fitted function
are shown in figure \ref{2076_paper_data} and \ref{2076_paper_fit}.

\begin{table}[h]
\caption{Fit results of the real data acquired by the GPD in the central position, in a frame of 120 $\mu$m $\times$ 170 $\mu$m around the peak position,
using the 2-D gaussian of equation \ref{bidimgauss}.}
\label{tab_res_fit}
\begin{center}
\begin{tabular}{|c|c|}
  \hline
  Param. & Fit results\\
  \hline
  $A$ & 95.36 $\pm$ 1.91 counts\\
  $X_\circ$ & (2.1 $\pm$ 0.4) $\mu$m \\
  $Y_\circ$ & (-17.4 $\pm$ 0.5) $\mu$m \\
  $\sigma_{X}$ & (29.5 $\pm$ 0.4) $\mu$m \\
  $\sigma_{Y}$ & (37.3 $\pm$ 0.4) $\mu$m \\
  \hline
\end{tabular}
\end{center}
\end{table}

Having determined in section \ref{sec: Beam_profile} the main parameters of the beam profile we are in the position to correct the measured values.
We subtracted quadratically the size of the beam as derived in table \ref{tab_res_beam_scan} from the
position resolution of the GPD as in table \ref{tab_res_fit}. The results for the two standard deviations are $\sigma^{GPD}_{X}$ = (28.1 $\pm$ 0.5) $\mu$m
and $\sigma^{GPD}_{Y}$ = (34.3 $\pm$ 0.6) $\mu$m.
The remaining significative, albeit small, difference in size between the two directions could be due to the different sampling in X and Y because of the
hexagonal readout pattern. While X is sampled with 50 $\mu$m resolution, thus the hexagonal pitch, Y is sampled with a resolution of 50 $\mu$m $\times$ cos($30^\circ$)
= 43.3 $\mu$m. However we cannot exclude, also, a residual misalignment of the X-ray beam axis with respect to the GPD plane and the effect on the different elongation
due to the inclined penetration. The control of such angle is, indeed, mostly relied on machined parts and an alignment procedure based on measurements at controlled
different inclinations will be studied. For the time-being we retain that $\sigma^{GPD}_{X}$ and $\sigma^{GPD}_{Y}$ are close enough to assess that no major
problems in the set-up, in the CMOS ASIC read-out or in the algorithm are present.

\section{The Monte Carlo simulation of the beam profile}
\label{sec: Monte_Carlo}
\begin{figure}
\centering
\includegraphics [scale=0.4]{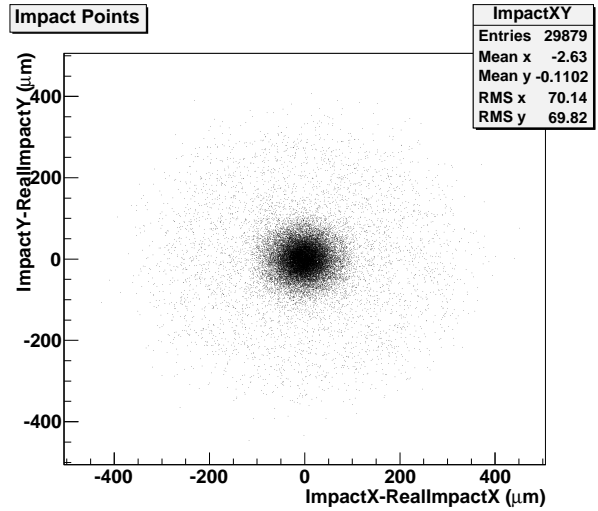}
\caption{Scatter plot of the reconstructed impact points from simulated data.
Wings are defined as the set of impact points outside a core of 200 $\mu$m of diameter.}
\label{Monte_scatter_XY}       
\end{figure}

A simulation was performed with the Monte Carlo we routinely use to
anticipate the polarimetric performances of the GPD. The Monte
Carlo simulates the generation, the transport, the multiplication
and the collection of the Auger-electrons tracks and of the photoelectron tracks from different atom
shells. Such tracks are generated by X-rays of a given polarization degree and of a given energy in a given gas
mixtures. The main routines of the Monte Carlo were earlier described
in \cite{2001NIMPA.469..164S, 2003SPIE.4843..383B, 2003SPIE.4843..394P}.
The simulated data are processed by algorithms equivalent to those used for the real data collected by the GPD.
The agreement between the Monte Carlo and the experimental data in terms of modulation factor
was already studied for different gas mixtures and detectors, see \cite{2008NIMPA.584..149M, 2010NIMPA.620..285M, 2012SPIE.8443 }.
Here, instead, we compare the position resolution.
At this regards we simulated a narrow pencil beam of 4.5 keV unpolarized photons impinging on the center of the GPD.
For each simulated track we derived the reconstructed impact point, together with the reconstructed emission direction.
The total number of counts in the simulation is larger with respect to the acquired real data of a factor of about 3.
The location of the impact points resulting from the simulation is shown in figure \ref{Monte_scatter_XY}. The presence of a core
and external wings, evident in this image, will be characterized in section \ref{subsec: core_wings} and
\ref{sensitivity_core_wings}. In this paragraph we show the results of the fitting procedure to determine the parameters of the simulated
image profile.

\begin{figure}[htpb]
\centering
\subfigure[\label{MonteCarlo_data}]{\includegraphics[scale=0.30]{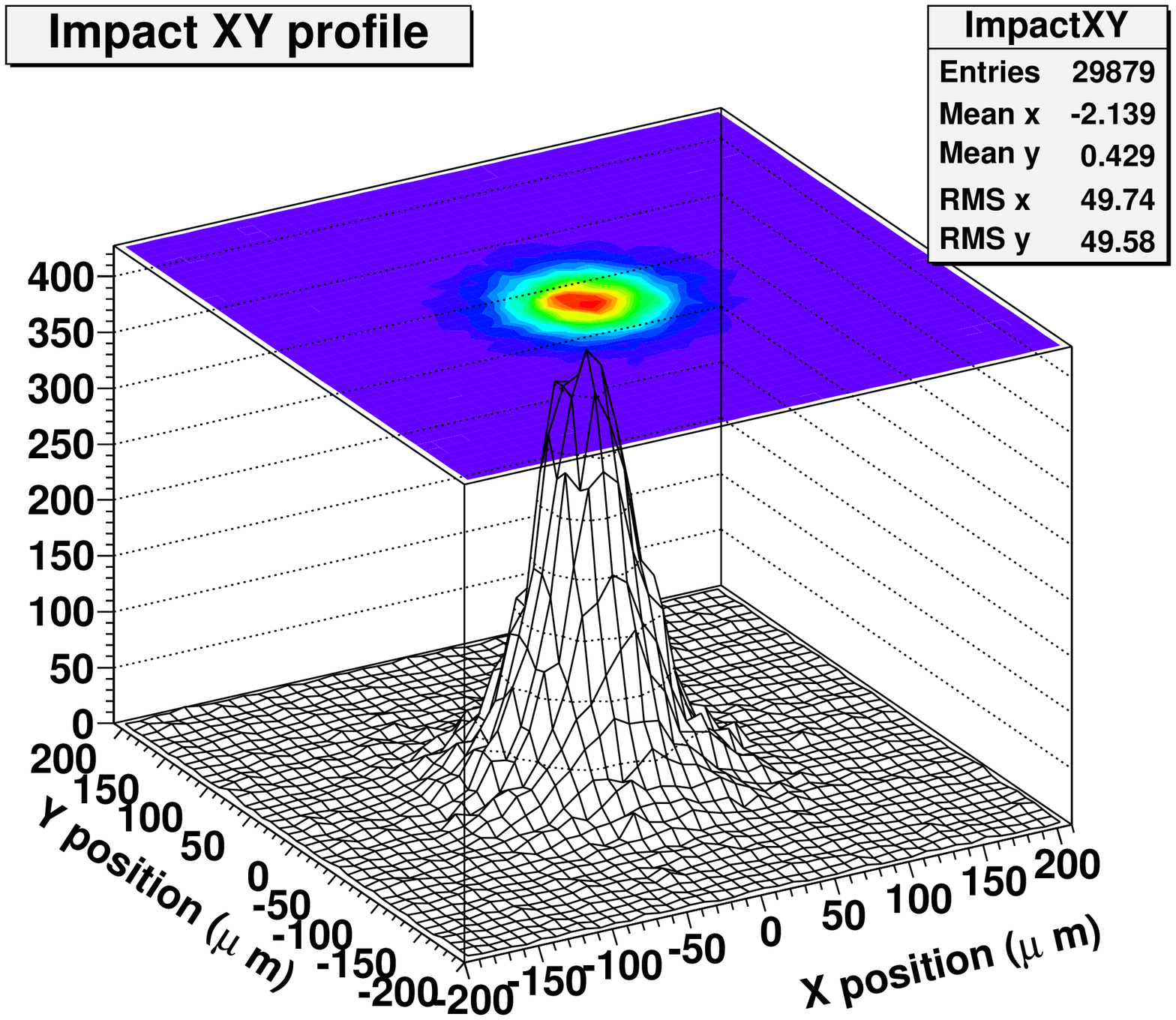}}
\hspace{0cm}
\subfigure[\label{MonteCarlo_fit}]{\includegraphics[scale=0.30]{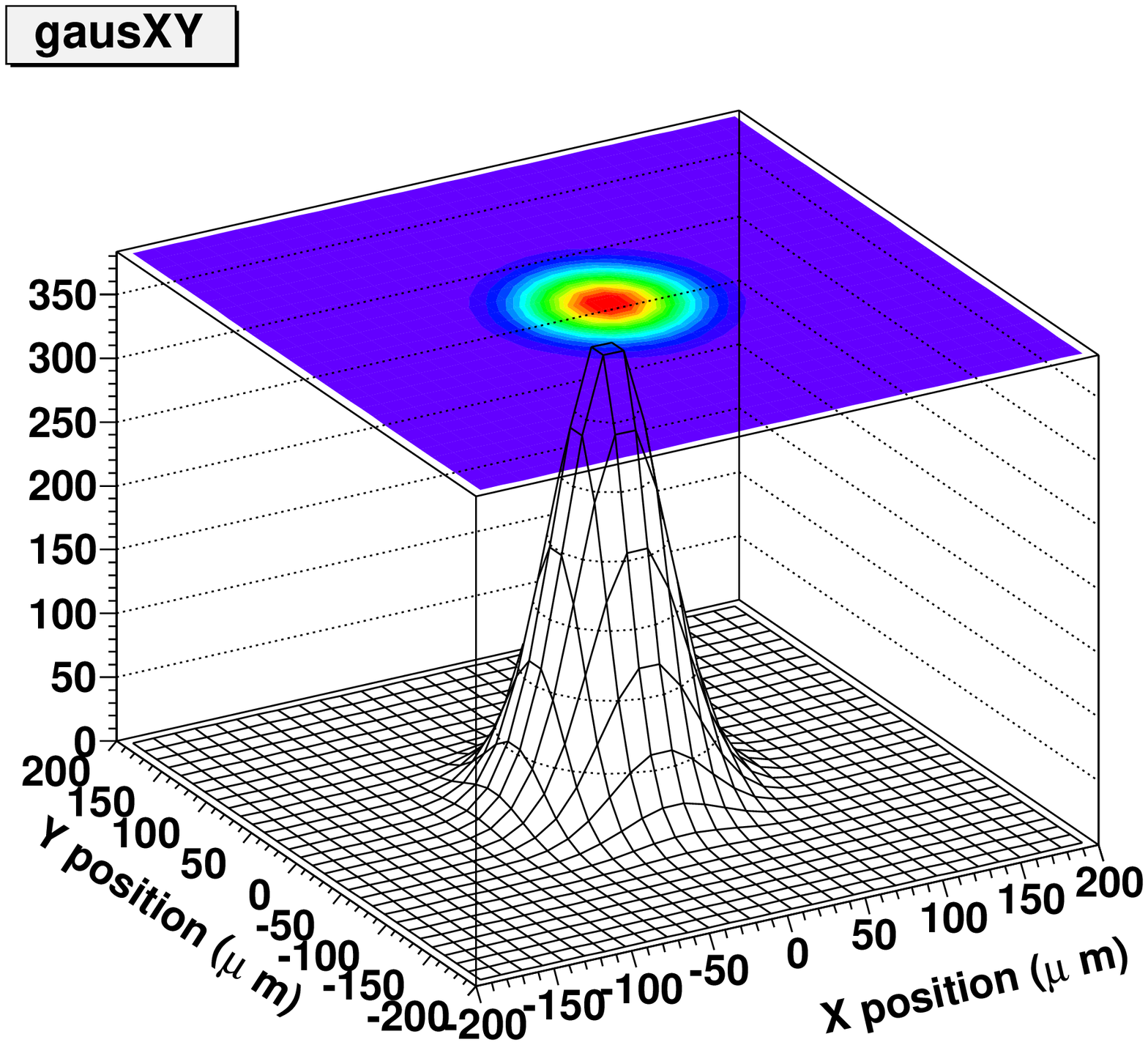}}
\caption{({\bf a}). Monte Carlo simulation of a narrow beam of 4.5 keV as imaged by the GPD. ({\bf b}). Gaussian fit of the core (within a square
wide 110 $\mu$m $\times$ 110 $\mu$m) of the simulated data.}
\end{figure}

Such image profile, together with the 2-D gaussian function used to fit the data (within a square wide 110 $\mu$m $\times$ 110 $\mu$m),
is, indeed, displayed in figure \ref{MonteCarlo_data} and \ref{MonteCarlo_fit} respectively.

\begin{table}[h]
\caption{Results of the fitting of the core of the simulated beam image with the 2-D gaussian function of equation \ref{bidimgauss}.}
\label{tab_Monte_Carlo_res_fit}
\begin{center}
\begin{tabular}{|c|c|}
  \hline
  Param. & Fit results\\
  \hline
  $A$ & 363.9 $\pm$ 4.4 counts\\
  $x_\circ$ & (0.00 $\pm$ 0.02) $\mu$m \\
  $y_\circ$ & (0.00 $\pm$ 0.08) $\mu$m \\
  $\sigma_{x}$ & (30.8 $\pm$ 0.3) $\mu$m \\
  $\sigma_{y}$ & (31.9 $\pm$ 0.3) $\mu$m \\
  \hline
\end{tabular}
\end{center}
\end{table}

The results of the fit are in table \ref{tab_Monte_Carlo_res_fit} and the
$\chi^{2}_{\nu}$ is now 2.2. Such a large value may be is due to the contribution of the wings which are included in the corner of the
selected region. Actually fitting the same function on a larger set of the data (a square wide of 130 $\mu$m $\times$ 130 $\mu$m), the
$\chi^{2}_{\nu}$ increases to 2.5 probably because of the larger contribution of the wings.

If we compare the results of table \ref{tab_Monte_Carlo_res_fit} with $\sigma^{GPD}_{X}$ and $\sigma^{GPD}_{Y}$,
we see that they agree within about 3 $\mu$m. The agreement is very good also considering that the simulated beam is aligned by construction
with respect to the GPD plane.

\subsection {The core and wings in the simulated image}
\label{subsec: core_wings}
Puzzled by the presence of the wings, we used the Monte Carlo data to investigate their origin and to study their contribution to the sensitivity
for polarimetry. With the Monte Carlo is, indeed, easier to relate the evolution of the original simulated track to the event detected by the GPD
after transport, blurring, multiplication and collection. The exact definition of the core and the wings is beyond the scope of this paper.
Therefore we define, somewhat arbitrarily, the core as the region
inside 200 $\mu$m in diameter from the peak position, and the wings as the set of the impact points outside the core. With this definition in mind
we compared the detected number of events with those expected by the 2-D symmetrical gaussian function. We recall that the density of probability
(equation \ref{bidimgauss}) with the normalization constant $\frac{1}{2\pi \times \sigma^{2}}$, if circularly symmetric, can be written as :

\begin{equation}
\label{bidimgauss_symmetric}
    f(r) =  \frac{1}{\sigma^{2}} \times r \times e^{- \frac{r^{2}}{2 \sigma^{2}}}
\end{equation}

The fraction of events outside a circle of radius $r_{0}$ can be written as in equation \ref{bidimgauss_symmetric_int} :

\begin{equation}
\label{bidimgauss_symmetric_int}
         \int_{r_0}^\infty f(r) dr =  e^{- \frac{r_{0}^2}{2 \sigma^2}}
\end{equation}

The fraction of the reconstructed impact points outside a radius of 100 $\mu$m in the Monte Carlo data
is indeed 16.6 $\%$, while the expected fraction is only 0.5 $\%$. This excess represents the contribution of the wings.
The investigation of the wings is important because some wings are, generally, present also in
the Lorentian PSF of the X-ray optics \cite{1994SPIE.2279..480C, 1996SPIE.2805...56C} and an interplay between the two could
be expected in focal plane experiments.

\begin{figure}[htpb]
\centering
\subfigure[\label{Traccia_Monte_orig}]{\includegraphics[scale=0.20]{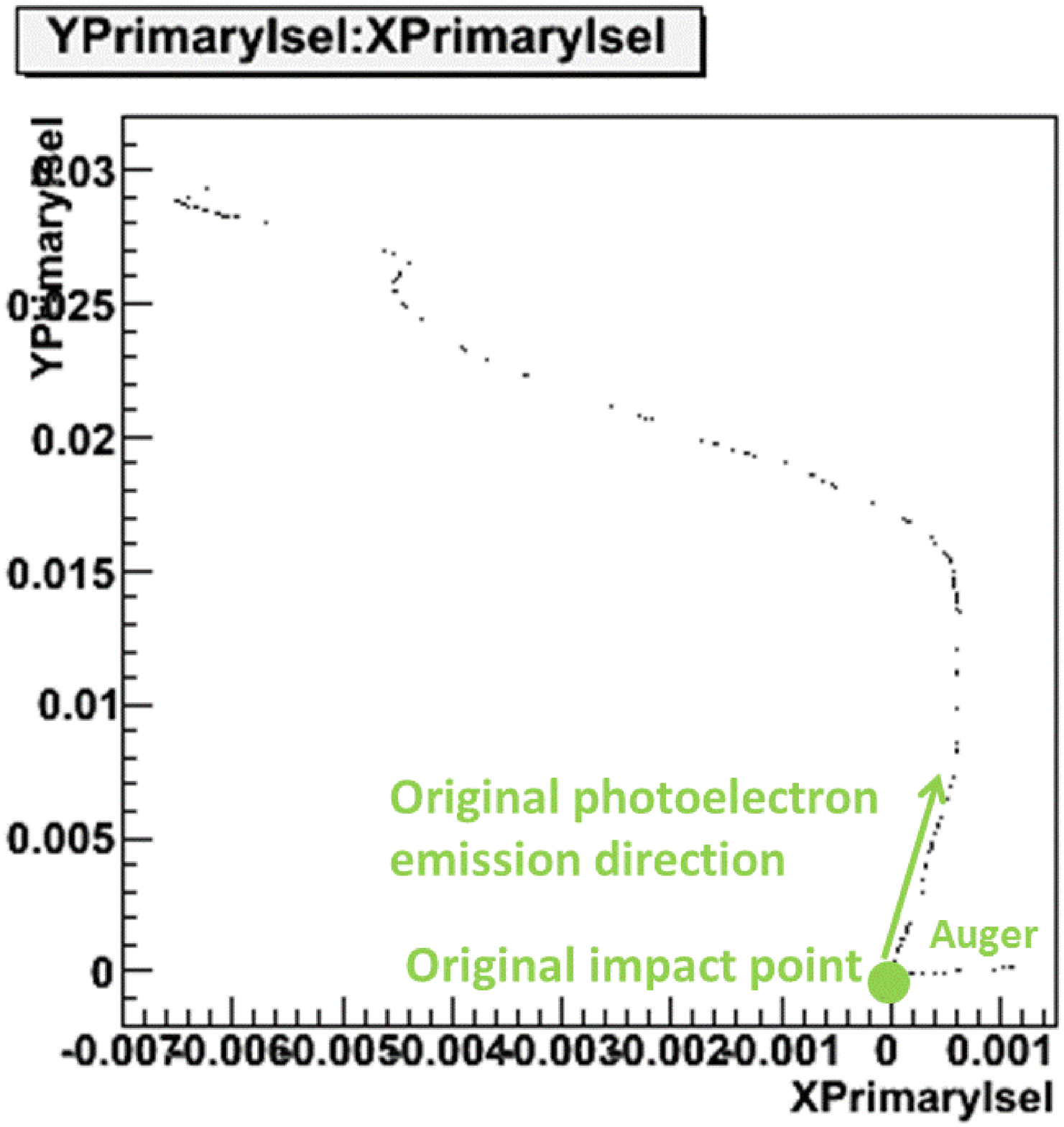}}
\hspace{0cm}
\subfigure[\label{Traccia_Monte}]{\includegraphics[scale=0.30]{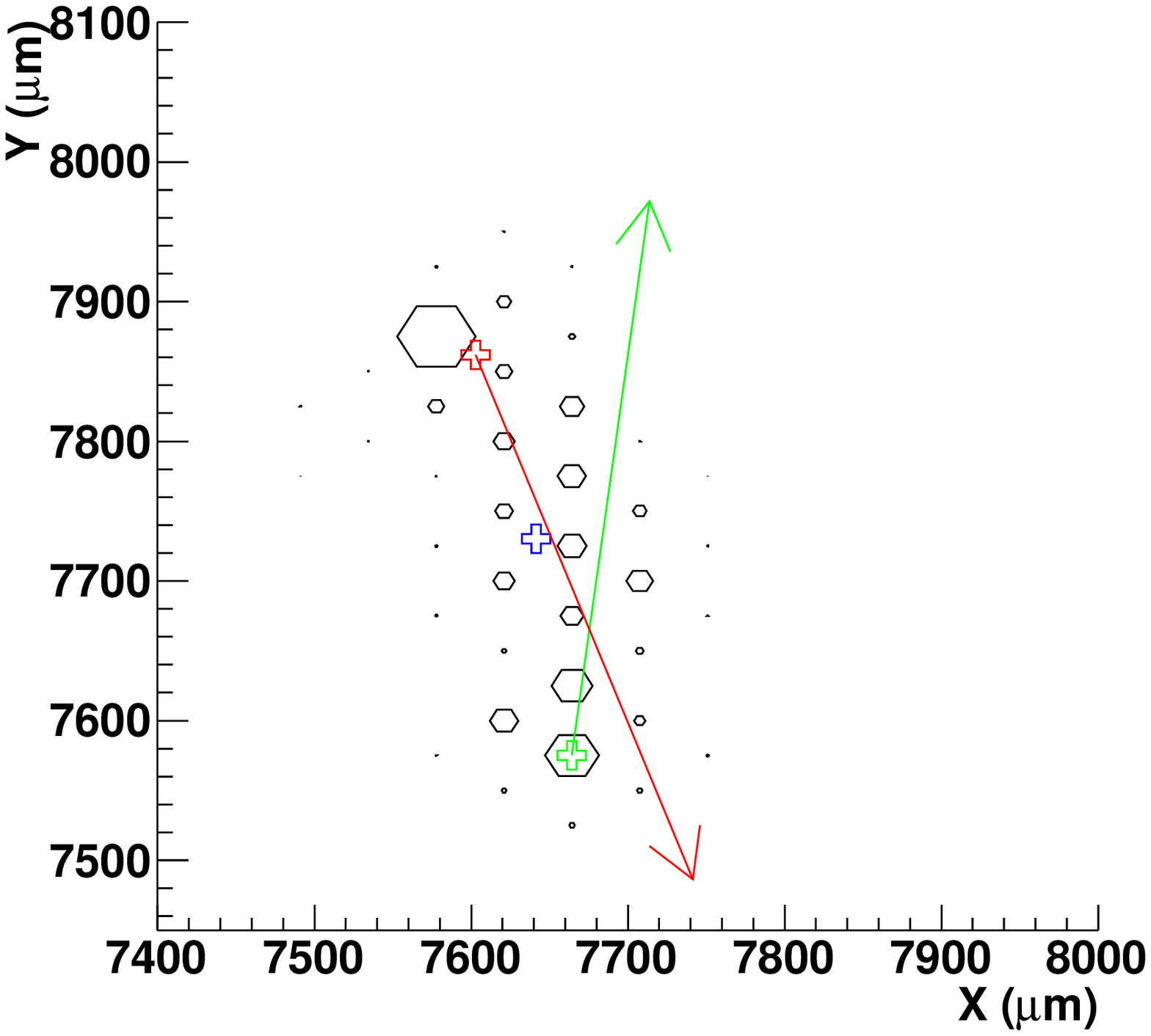}}
\caption{({\bf a}). A simulated track in the wings projected on the GPD plane. ({\bf b}). The same simulated track after transport, blurring,
multiplication and collection. The green cross and the arrow line is the original impact point with the emission direction while the red cross and arrow line
is the reconstructed impact point and the reconstructed emission direction.}
\end{figure}

An inspection of the Monte Carlo tracks with calculated positions outside the core showed that the presence of wings
is due, indeed, to an incorrect determination of the reconstructed impact point. In practice the reconstructed impact point
is on the opposite side with respect to real impact point. This happens because the algorithm calculates the skewness
for determining the end point of the track (see section \ref{sec: GPD&algorithms}) containing the Bragg peak that is the end of the photoelectron evolution.
The distribution of charges for some kind of trajectories provides blurred collected tracks with the skewness
giving the '\emph{wrong}' sign. Consequently, instead to identify in the track the correct end point,
the incorrect one is derived. A simulated track where this inversion happens is shown in figure \ref{Traccia_Monte_orig}
where it is displayed its projection on the X $\&$ Y plane (the GPD readout plane) in the Monte Carlo coordinate system.
In figure \ref{Traccia_Monte} it is shown the image of the same track blurred by diffusion,
multiplied by the GEM and collected by the readout plane. The green cross, in the latter, is the original impact point,
the blue cross is the barycentre and, finally, the red cross is the reconstructed position.
An inspection of the original track shows, see figure \ref{Traccia_Monte_orig}, that
the impact point in the original frame coordinates is at (X = 0; Y = 0), the Auger-electron evolves rightwards while the photoelectron
firstly evolves in the same semi-plane containing the Auger-electron, secondly scatters on a nuclei and moves on the opposite
semi-plane. The derived distribution of the blurred charge is such that a larger density is at the end point containing the original impact point and
not at the opposite end point containing the Bragg peak: the evaluation of the skewness, therefore, produces an inversion.
A consequence of such an error is that the emission direction is, also, changed (see again figure \ref{Traccia_Monte} and its caption)
and almost inverted. Due to the $180^\circ$ symmetry a certain, albeit smaller, sensitivity to polarization is expected to
be preserved also in the events contained in the wings. This is the object of the study presented in the section \ref{sensitivity_core_wings}.

\subsection {The core and wings: the impact on the polarization sensitivity}
\label{sensitivity_core_wings}
We studied  also by Monte Carlo simulations the contribution of the core and wings to the polarimetric sensitivity of the GPD.
Such study, in our laboratory, at the present time can only be done by simulation. The generation of polarized radiation,
in our facility, greatly reduces the flux from the X-ray tube and, therefore, the use of a collimator system with a narrow beam in addition
to the polarizer is of no practical use.

However the agreement between the Monte Carlo data and the experimental data, both in terms of modulation factor and position resolution,
is such that a comparative study of the wings and the core in terms of polarization sensitivity is reasonable in this way.

\begin{figure}[htpb]
\centering
\subfigure[\label{Mod_core}]{\includegraphics[scale=0.3]{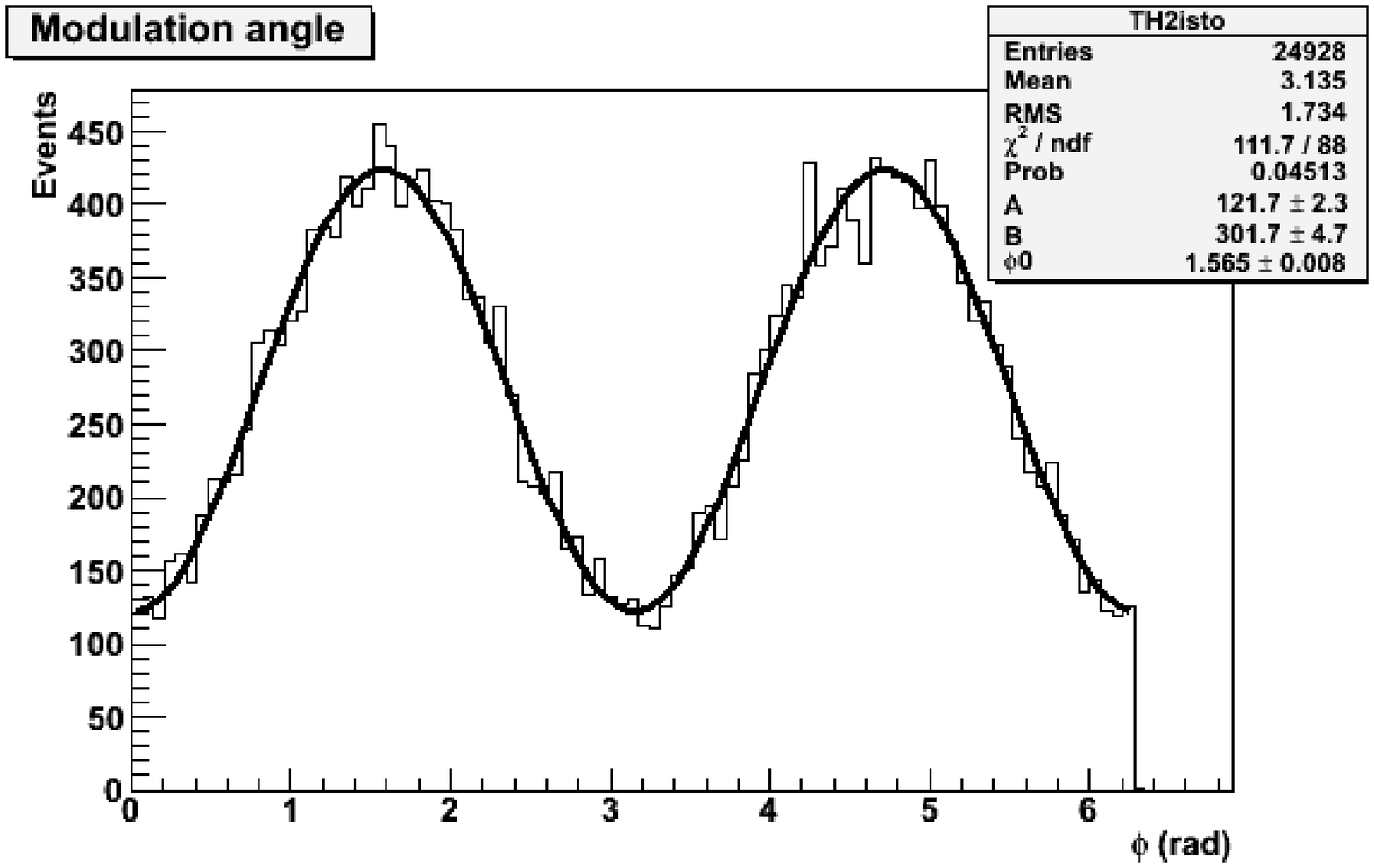}}
\hspace{0cm}
\subfigure[\label{Mod_wings}]{\includegraphics[scale=0.3]{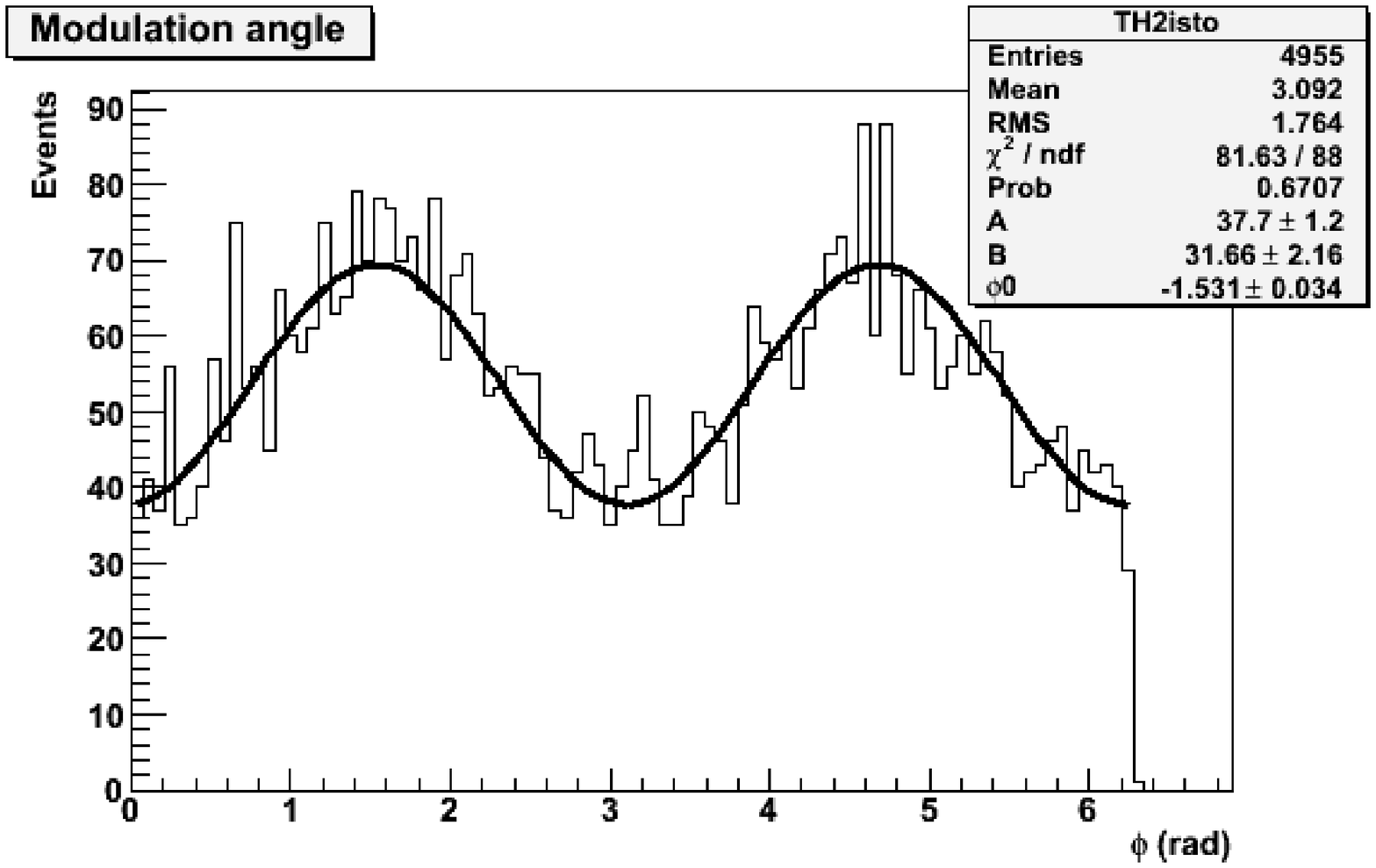}}
\caption{({\bf a}). Modulation curve from simulated $100\%$ polarized 4.5-keV photons in the core. The modulation
factor is $\mu$  = (55.3 $\pm$ 0.6) $\%$. ({\bf b}). Modulation curve from simulated $100\%$ polarized 4.5-keV photons in the wings. The modulation
factor is $\mu$  = (29.6 $\pm$ 1.6) $\%$.}
\end{figure}

We simulated for this study a zero-width narrow X-ray beam of 4.5 keV and 100 $\%$ polarized and we derived the reconstructed impact point and the
emission direction for each event.
The histogram of the emission directions, the so called modulation curve, is reported in figure \ref{Mod_core} and in figure \ref{Mod_wings}
for the core and the wings respectively with the following function to derive the modulation factor :

\begin{equation}
    f(\phi) = A + B   cos^{2}(\phi- \phi_0)
\label{cos2}
\end{equation}

where A and B and $\phi_0$ are parameters of the fit.
The modulation factor $\mu$ is given by the equation \ref{mu} :

\begin{equation}
\label{mu}
    \mu = \frac{B}{2A+B}
\end{equation}

The modulation factor of the whole data set is $\mu$ = (50.9 $\pm$ 0.6) $\%$. The contribution of the core is
$\mu$  = (55.3 $\pm$ 0.6) $\%$ and the contribution of the wings decreases down to $\mu$  = (29.6 $\pm$ 1.6) $\%$. We therefore
demonstrated that the core, as expected, has a larger polarization sensitivity with respect to the wings. This is interesting because,
by selecting the events contained in the core, we could determine the polarization of a point-like celestial source with a
better confidence. We recall, actually, that the polarization sensitivity is proportional to the so called quality factor (equation \ref{QF}) :

\begin{equation}
\label{QF}
    QF = \mu\sqrt{\varepsilon}
\end{equation}

where $\varepsilon$ is the efficiency. In our case $QF_{core}$ = 0.505 and $QF_{all}$ = 0.509 therefore basically identical, however
the larger modulation factor obtainable selecting only the events in the core allows for a slightly better control of the incidence on the measurement
of possible residual systematics. Moreover the lower, but not zero, modulation factor of the wings allow for using these events once
properly weighted. In any case the analysis shows that the main origin of the wings is that, for a fraction of events, mainly those characterized
by photoelectrons that underwent to a large scattering angle with an atomic nucleus, the skewness is not sufficient to identify the section of
the track that includes the impact point. This means that an improved pattern recognition can be mitigates significantly the problem.

The effects on sensitivity of the selection of the core and the wings needs a more accurate study. For example they are expected to be
energy dependent through the different average size and shape of the tracks. Such study is beyond the scope of this work, but
we recall that spectral capability of the GPD allows for such an investigation.

\section{The effect of the large guard-ring on the image profile}

\begin{figure}
\centering
\includegraphics [scale=0.4]{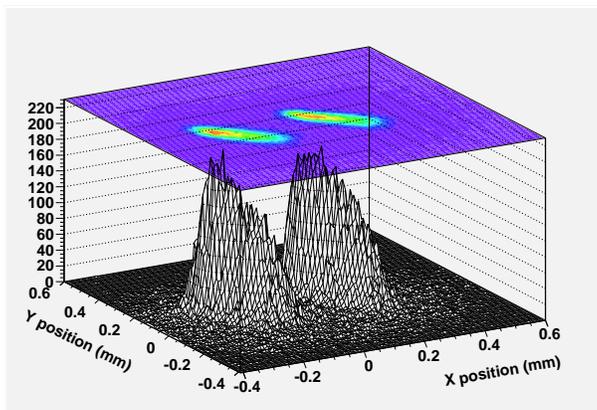}
\caption{ Distortion of the two images of the beam shifted 300 $\mu$m after the disconnection of the upper guard-ring.}
\label{guard_ring_effect}       
\end{figure}

The two different configurations up to now used for the GPD, the old configuration, with a small guard-ring in the GEM plane,
and the present configuration, with a larger guard-ring, were already described in section \ref{sec: GPD&algorithms}.
We can not directly compare the two configurations in terms of position resolution, since the
old configuration is not available at the moment, but we can show the effect of the presence of the guard-ring on the location capability of the detector.
Therefore, to study how the uniformity of the electric field impacts on the image of the narrow beam, we disconnected the upper guard-ring of the
GEM plane.
The results are shown in fig. \ref{guard_ring_effect} for two of the three positions of fig. \ref{3_beams}. The effect
is indeed enormous. The image of the beam  appear now larger, distorted and rotated even at the center of the active region at about 7.5 mm
from its border.

The electric-field, with this disconnection, is not the same of that of the old-one, with a smaller guard-ring and a smaller aspect ratio (the ratio
between the thickness of the drift region and the linear dimension of the surface area of the GEM plane). However this measurement suggests
that the present configuration assures better performances in terms of drift-field uniformity and position resolution capability.

\section{Conclusions}
\label{Conclusions.}
We studied the position resolution of the GPD by means of laboratory measurements and Monte Carlo simulations, at 4.5 keV, using a narrow X-ray beam. After
subtracting the contribution of the beam size which was measured by means of a scanning technique, we obtained a position resolution
more than twice better than that already found with a Micro Pattern Gas Detector with 200 $\mu$m pitch and a GEM with a larger pitch
\cite{2003SPIE.4843..383B}.
The measured position resolution is very close to the Monte Carlo results. By using the latter and equation \ref{bidimgauss_symmetric_int},
the Half Energy Width (HEW, spatial) found is 36.3 $\mu$m. The expected HEW  (spatial) by the JET-X optics
(with a focal length of 3500 mm and a HEW (angular) of 19.3'' \cite{2010xpnw.book...79L}), is 327 $\mu$m, a factor of nine larger,
including also the effects of inclined penetration of the photons collected in the absorption depth of the GPD.
The angular resolution of an experiment with the GPD at the focus of such X-ray optics is, therefore, driven primarily by the quality of the latter.

We detected the presence of wings in the image of the beam in both the real data and the Monte Carlo data.
Such wings were studied primarily by simulations and are due to an inversion in the location
of the impact point due to peculiar event trajectories and to the blurring. The core events have a larger modulation factor with respect to the events in the
wings suggesting that a spatial selection of the data, a different weight, or a better pattern recognition,
especially for polarimetry of point-like celestial sources, would lead to an improved statistical significance.

Finally we showed the effect of the presence of the large guard-ring on the image profile of the narrow beam.







\section{Acknowledgment}
This work is partially supported by the Italian Space Agency (ASI).

\end{document}